\documentclass[10pt,a4paper]{ouparticle} 
\usepackage[inline]{enumitem}
\usepackage{siunitx}  
\usepackage[backend=biber,      style=nejm,
doi=false,                  url=false,                  isbn=false,                 natbib=true,
]{biblatex}
\usepackage{hyperref,eurosym,multirow,xcolor,colortbl}
\usepackage{cleveref,tabularx}
\usepackage[normalem]{ulem}
\usepackage{xspace}
\usepackage{array}
\usepackage{paralist}
\usepackage{wasysym} 
\usepackage{lineno}
\usepackage[font={sf,footnotesize},labelfont=bf]{caption}
\usepackage[T1]{fontenc}

\newcommand{\rtwo}{0.5}

\newcommand{\pone}[1]{\includegraphics[scale=#1]{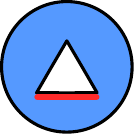}}
\newcommand{\ptwo}[1]{\includegraphics[scale=#1]{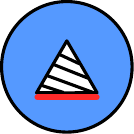}}
\newcommand{\pthree}[1]{\includegraphics[scale=#1]{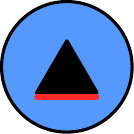}}
\newcommand{\pfour}[1]{\includegraphics[scale=#1]{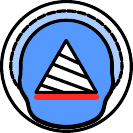}}
\newcommand{\pfive}[1]{\includegraphics[scale=#1]{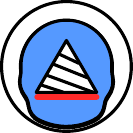}}
\newcommand{\psix}[1]{\includegraphics[scale=#1]{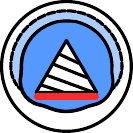}}
\newcommand{\pseven}[1]{\includegraphics[scale=#1]{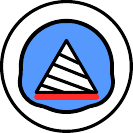}}

\newcommand{\cnum}[1]{{\textcircled{\footnotesize {#1}}}}
\newcommand{\pA}{entrainment phase} 
\newcommand{\PA}{Entrainment phase} 
\newcommand{\pB}{steady phase} 
\newcommand{\PB}{Steady phase} 
\newcommand{\pC}{damping phase} 
\newcommand{\PC}{Damping phase} 
\newcommand{\pp}{peak-to-peak amplitude}

\newcommand{\ppa}{peak-to-peak amplitude}
\newcommand{\scg}{semi-circular grooves}
\newcommand{\grooves}{semi-circular grooves}
\newcommand{\Grooves}{Semi-circular grooves}
\newcommand{\groove}{semi-circular groove}
\newcommand{\scgrooves}{semicircular grooves}
\newcommand{\dip}{ventral dip}
\newcommand{\dln}{dentate ligament network}
\newcommand{\dl}{dentate ligament}

\newcommand{\ligaments}{ligaments}

\newcommand{\canal}{spinal canal}

\newcommand{\Gb}{Glycogen body}
\newcommand{\gb}{glycogen body}
\newcommand{\cord}{spinal cord}
\newcommand{\Cord}{Spinal cord}

\newcommand{\als}{locomotion simulator}

\newcommand{\dof}{damped oscillation frequency}
\newcommand{\dofs}{damped oscillation frequencies}

\newcommand{\Dofs}{Damped oscillation frequencies}
\newcommand{\st}{settling time}

\newcommand{\fluid}{spinal fluid}
\newcommand{\Fluid}{Spinal fluid}
\newcommand{\flow}{fluid flow}

\newcommand{\bpm}{biophysical model}
\newcommand{\Bpm}{Biophysical model}

\newcommand{\spindle}{spindle}
\newcommand{\frequency}{drive frequency}
\newcommand{\frequencies}{drive frequencies}
\newcommand{\Frequencies}{Drive frequencies}
\newcommand{\lobes}{accessory lobes}

\newcommand{\Lobes}{Accessory lobes}
\newcommand{\rate}{decay rate}
\newcommand{\Rate}{Decay rate}
\newcommand{\lso}{lumbosacral organ}

\newcommand{\scord}{spinal cord}

\newcommand{\vs}{vestibular system}

\newcommand{\scanal}{spinal canal}

\newcommand{\msd}{mass-spring-damper}

\begin{document}
\title{Biophysical Simulation Reveals the Mechanics of the Avian Lumbosacral Organ}
\author{\name{An Mo$^1$, 
Viktoriia Kamska$^1$, 
Fernanda Bribiesca-Contreras$^1$, 
Janet Hauptmann$^{1,3}$, 
Monica Daley$^2$, 
Alexander Badri-Spr\"owitz$^{1,4}$}
\address{$^1$Dynamic Locomotion Group, Max Planck Institute for Intelligent Systems, Stuttgart, Germany}
\address{$^2$Department of Ecology and Evolutionary Biology, University of California, Irvine, USA}
\address{$^3$Harz University for Applied Sciences, Wernigerode, Germany}
\address{$^4$Department of Mechanical Engineering, KU Leuven, Leuven, Belgium}}
\date{}
\maketitle \begin{abstract}{}
The lumbosacral organ (LSO) is a lumbosacral spinal canal morphology that is universally and uniquely found in birds. Recent studies suggested an intraspinal mechanosensor function that relies on the compliant motion of soft tissue in the spinal cord fluid. It has not yet been possible to observe LSO soft tissue motion in vivo due to limitations of imaging technologies. As an alternative approach, we developed an artificial biophysical model of the LSO, and characterized the dynamic responses of this model when entrained by external motion. The parametric model incorporates morphological and material properties of the LSO. We varied the model's parameters to study the influence of individual features on the system response. We characterized the system in a locomotion simulator, producing vertical oscillations similar to the trunk motions. We show how morphological and material properties effectively shape the system's oscillation characteristics. We conclude that external oscillations could entrain the soft tissue of the intraspinal lumbosacral organ during locomotion, consistent with recently proposed sensing mechanisms.


\end{abstract}	

\section{Introduction} 
Birds are exceptional bipedal runners capable of robust running over unexpected disturbance~\citep{daley2006running}.
Robust locomotion requires a sense that informs the central nervous system about the environment and the system's internal state. 
Such sensing is essential to coordinate limbs~\citep{knuesel_effects_2011,conway_proprioceptive_1987}, balance~\citep{mouel_anticipatory_2019}, manipulate the environment~\citep{kuchenbecker_verrotouch_2010}, for entrainments~\citep{goldfield_infant_1993,taga_emergence_1994,berthouze_assembly_2008,ruppert_learning_2021}, and protect from excessive loading or untimely muscle stretching~\citep{haen_whitmer_mixed_2021}. 
Rapid sensing and response is crucial, especially during fast locomotion.
When stance phases are brief, a sensorimotor delay~\citep{more2018scaling} will cause a temporal blind spot in the control loop, potentially leading to detrimental falls.
Interestingly, birds generally have long necks, contributing to increased sensorimotor delays from higher brain centers, as well as from the balance-sensing vestibular system~\citep{urbina2018physical}.
The immediate physical response of the musculoskeletal~\citep{daley2009role} system alone cannot fully explain birds' agility.

\begin{figure*}[htb]
\centering
\includegraphics{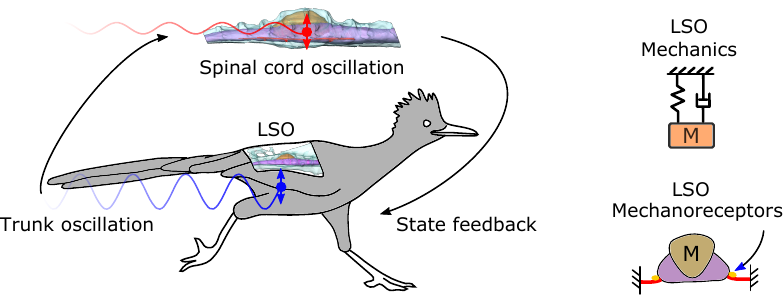}
\caption{The LSO located at the lower spine of birds is hypothesized as an accelerometer. 
During locomotion such as running, the truck oscillation will entrain the spine cord like a mass-spring-damper system.
The morphology of LSO tunes the mass-spring-damper behavior.
The entrained LSO stimulates the mechanoreceptors to provide fast state feedback of locomotion.
}
\label{fig:overview}
\end{figure*}

Birds' outstanding locomotion abilities might be supported by an unexplored and uniquely avian intraspinal mechanosensor: the \lso (LSO, \Cref{fig:overview})~\citep{necker_specializations_1999}.
It has been suggested that the LSO could act as a second vestibular-like sensing organ, independent of the head's orientation~\citep{necker_specializations_1999}. 
The LSO is located at the lower spine, right next to the sciatic nerves that communicate motor commands for locomotion~\citep{bekoff_coordinated_1975}.
The short distance between intraspinal mechanosensors and spinal motor-control units could minimize sensorimotor delays, and effectively reduce response times~\citep{kamska20203d}.

The LSO is a collection of unique anatomical features~(\Cref{fig:phantom}A).
A \gb{} is dorsally wedged between both \scord{} hemispheres, spanning over three segments~\citep{kamska20203d}.
At the LSO's centre, the \gb{} ventrally reaches the central canal~\citep{de_gennaro_ultrastructural_1976,moller_blood_brain_2003}.
\Lobes{} (``Hofmann nuclei'' or ``major marginal nuclei'', \citep{eide_axonal_1996}) are found pairwise, segmentally, and laterally to the lateral side of the \scord{}.
Potentially, they contain mechanoreceptors~\citep{schroeder_specializations_1987,necker_specializations_1999,Rosenberg_Necker2002,necker2006specializations,yamanaka2008chick,stanchak_balance_2020,stanchak_molecular_2022}.
Hoffmann nuclei processes project into ipsilateral and contralateral hemispheres~\citep{eide_axonal_1996,eide_development_1996}.
The \scord{} is supported ventrally by a complex \dln{}, comprised of lateral longitudinal, ventral longitudinal, and transverse ligaments~\citep{kamska20203d}.
The vertebrae in the LSO region are fused with fusion zones formed as transverse semi-circular grooves (``semi-circular canals''~\citep{necker2006specializations}).
Between the LSO soft tissue (\scord{}, \gb{}, \dln{}, \lobes{}) and the \scanal{} walls exists a significant fluid space with a prominent dip ventral to the LSO central region~\citep{kamska20203d}.

Since its first discovery in 1811~\citep{emmert1811beobachtungen}, the exact function of the LSO remains an enigma.
Early research suggested metabolic energy supply and myelin synthesis as potential functions for the \gb{} and \lobes{}~\citep{de_gennaro_ultrastructural_1976,benzo_glycogen_1981,benzo_hypothesis_1983}.
Schroeder, Murray and Eide~\citep{schroeder_specializations_1987,eide_axonal_1996} were the first to propose a mechanoreceptive function. 
They had found mechanoreceptor-like tissue in \lobes{} and therefore theorized that \dl{} strain is transmitted to and sensed by the adjacent \lobes{}.
Later, \citeauthor{necker_specializations_1999} proposed that \scgrooves{} and \fluid{} are integral parts of the sensor organ's function.
He hypothesized that \lso{} spinal fluid flow could excite mechanoreceptive \lobes{}~\citep{necker_specializations_1999,yamanaka_analysis_2012,yamanaka_glutamate_2013}.
Besides, he was the first to point to morphological similarities between the \lso{} and the \vs{}~\citep{necker_specializations_1999}. 
Otherwise, possible mechanical functions of the LSO are largely unexplored.
While conclusive evidence for the LSO sensing function is still missing, intraspinal mechanosensing has been found in a few animals; lampreys~\citep{grillner_edge_1984,mcclellan1993mechanosensory}, zebrafish~\citep{bohm_csfcontacting_2016,picton_spinal_2021}, and potentially in reptiles~\citep{schroeder_marginal_1986}.
Despite the difference between these animals and birds, the similar location of their intraspinal mechanosensors is intriguing~\citep{grillner_edge_1984,viana_di_prisco_synaptic_1990,picton_spinal_2021,stanchak_molecular_2022}, and suggests a homologous connection.

Based on our own observation of morphologies and material properties~\citep{kamska20203d}, we hypothesize a locomotion state sensing function of the LSO (\Cref{fig:overview}).
We suspect that the viscoelastic properties of the \scord{} and ligaments allow these structures to physically deflect and oscillate within the enlarged fluid space~\citep{kamska20203d}.
During locomotion, the truck oscillation, such as pitching, will entrain the \scord{} oscillation.
The resulting soft tissue motion could resemble a \msd{} system; the dense \gb{} as the mass, the elasticity of the \scord{} and the \dln{} as the spring, and the \fluid{} as the damper.
The relative motion between the \scord{} and the \scanal{} would stretch the mechanoreceptors contained in the \lobes{}, then accelerations and postural changes could be measured, leading to a fast state feedback of locomotion.

In this work, we focus on the \msd{} properties of the LSO.
Since birds feature the highest number of locomotion modes within species; they swim, dive, walk, and fly; their habitats and locomotion modalities may shape the LSO response through \msd{} property variation.
Structures like the \gb{} with densities higher than the \fluid{} will tend to sink, exerting forces on the \dln{}.
The \gb{} are subject to growth~\citep{watterson_development_1949}, allowing for lifelong tuning and adaptation.
The microfluidic environment of the \scanal{} implies an effective flow resistance (Hagen–Poiseuille equation) to dampen high-frequency oscillation, similar to a mechanical low-pass filter.
Neural tissue is fragile, with a reported maximum strain up to \SI{7}{\%} for uni-axial fibre strain~\citep{tamura_variation_2007}.
Likely, the combined structure of ligaments, \fluid{}, and \gb{} protects the \scord{} from excessive strain.

While the spinal soft tissue entrainment is likely, observing such entrainment within a running bird is a grand challenge.
In birds, the \scord{} is well protected within the dense, fused bone structure.
Imaging the soft tissue motion in vivo has failed so far.
As an alternative approach, we developed a parametric, biophysical \lso{} model, which we based on previously reported data~\citep{kamska20203d}.
In sum, we suggest three hypotheses related to the \msd{} properties of the LSO.
\begin{inparaenum}[1)]
  \item The \gb{} tunes the LSO measurement range.
  \item The narrow \scanal{} dampens soft tissue oscillation.
  \item The fine structure of the \scanal{} diversifies the LSO response.
\end{inparaenum}



\section{Materials and Methods}

\begin{figure*}[htb]
  \centering
  \includegraphics[width=0.7\textwidth]{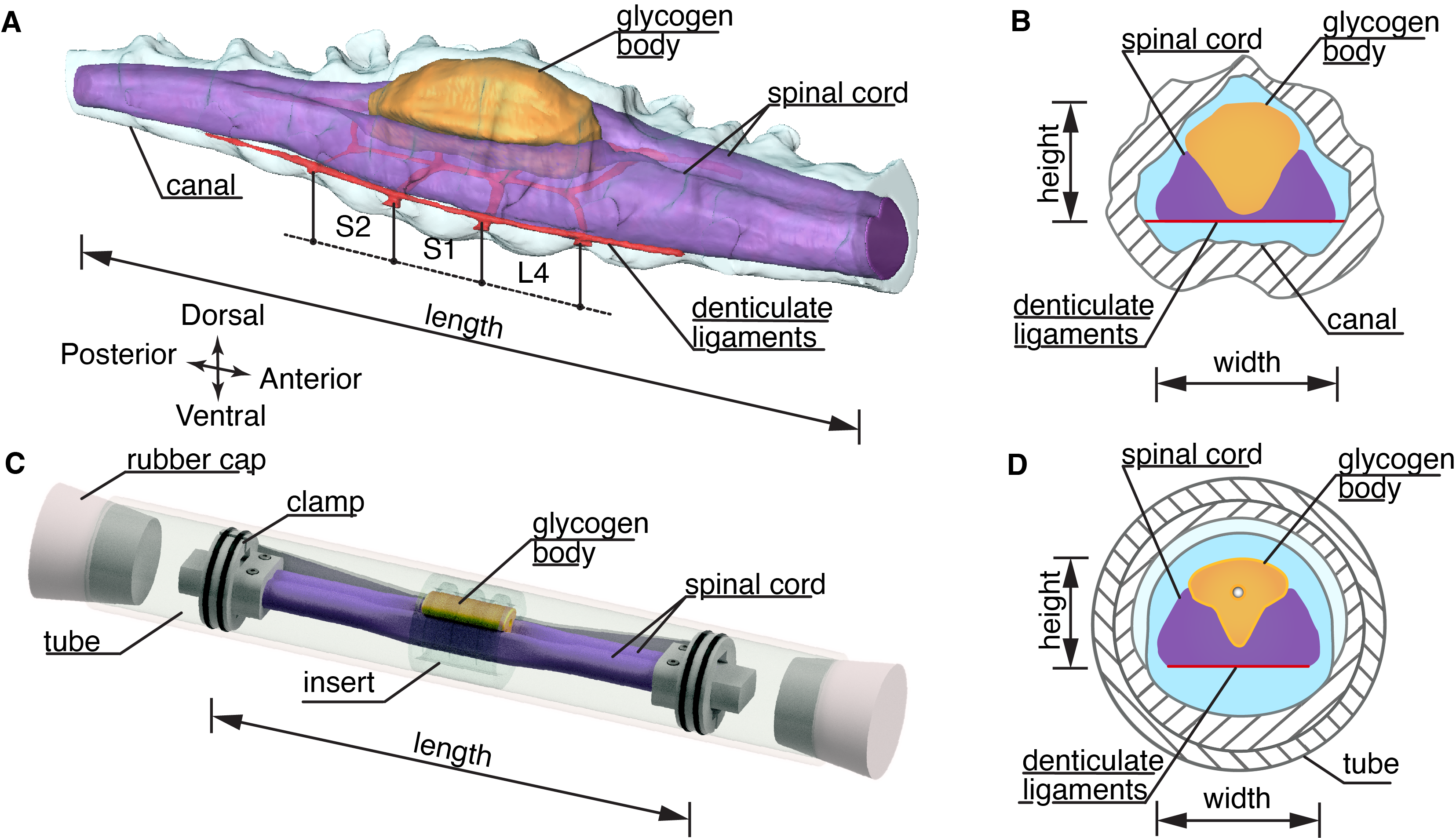}
  \caption{Lumbosacral soft tissue in the \canal{}, and its \bpm{} developed here to simulate soft tissue motion. 
  (A) Perspective view of the lumbosacral region of a common quail~\citep{kamska20203d}. Shown are \cord{} (purple), \gb{} (orange), and \dln{} (red). 
  (B) Transverse section through the vertebral column at vertebra fusion S1 and L4. 
  (C) \Bpm{} with \cord{}, \gb{} and \grooves{} mounted as a water-filled glass tube. In this configuration, the modelled \cord{} is clamped at its both ends. Transverse \scg{} are cut into the insert as indentations.
  (D) Cross-cut view of the \bpm{} at the position of a \groove{}.
  }
  \label{fig:phantom}
\end{figure*}

First, We developed a configurable \bpm{} of the lumbosacral organ in birds (\Cref{fig:phantom}).
We parameterized the \bpm{}’s morphology and varied its material properties to investigate the individual influence of each part and its associated hypothesis (\Cref{tab:test_plan}).
Then, the \bpm{}s were tested on a custom-built locomotion simulator (\Cref{fig:oscillator}), which emulates vertical locomotion patterns of running birds.
Lastly, the \bpm{}’s response to external accelerations was recorded and characterized (\Cref{fig:time_response_example}).

\subsection{Biophysical model}

We aimed at these goals to implement and test the \bpm{} of birds' LSO:
\begin{inparaenum}
\item Develop a simplified, parametric model for physical testing, with a low, appropriate number of design parameters. In contrast, a one-to-one replicated LSO geometry would lead to a large parameter number, which is infeasible for physical testing.
\item Select model parameters according to their relevance for the LSO's physical functionality according to our hypotheses.
\item Create an LSO model of appropriate size for fabrication and instrumentation. 
\item The ratio of volumes, material densities, and soft material stiffness approximates to data from the literature.
\end{inparaenum}

To replicate the geometry, we simplified and linearly scaled up the three-dimensional common quail model \textit{(Coturnix coturnix)} made available by Kamska et al.~\citep{kamska20203d}.
Its main components were simplified as in the \bpm{}: an \cord{}, a \dln{}, a \gb{}, the surrounding \fluid{}, the \canal{} morphology  between spine segments L4 to S2 (\Cref{fig:phantom}).
We linearly scaled up the LSO soft parts, leading to model parameters documented in \Cref{tab:lso_scaling}.
The length of the \bpm{} is \SI{140}{mm} between proximal and distal anchor points (\Cref{fig:phantom}C), which is roughly the size of the lumbosacral region of an emu~\citep{bausch_spezialisierungen_2014}.
We kept the volume ratio constant for the \cord{}, the \gb{}, and the \fluid{} (\Cref{tab:lso_scaling}).
We implemented model morphologies mimicking dorsal grooves and a ventral dip found in birds; both features were volume scaled.

To approximate the material properties, we fabricated the \bpm{} with soft robotics techniques.
The \cord{} and the \gb{} were moulded from silicone rubber, with \dln{} made from fabric attached.
The \gb{} density is adjustable.
A custom clamp holds the spinal soft tissue in a water-filled glass tube, simulating the fluid environment.
A configurable insert existed in some \bpm{}s to implement the \canal{} morphologies.
Detail fabrication steps are provided in the supplementary section S1. 
In sum, we prepared seven configurations of the \bpm{} as shown in \Cref{tab:test_plan}.

The resulting \bpm{} allows for characterizing its compliant parts responding to external motions while interacting with the surrounding fluid and complex canal morphologies.

\begin{table}[htb]
\small \caption{\Bpm{} design parameters. Volume percentages of the \bpm{} are in reference to the quail model~\citep{kamska20203d}, for the sum of volumes at L4-S2 region.}
\label{tab:lso_scaling}
\begin{center}
\begin{tabular}{llll}
Unit								& Parameter			 			& Reference		& Design \\
\hline
\multirow{3}{4em}{Length [\si{mm}]}	& Width, 			 $w$ 		& $ 3.5$ 		& 21  \\
									& Length,			 $l$ 		& $ 20$ 		& 140 \\
									& Height, 			 $h$ 		& $ 5.0$		& 30  \\
\hline
\multirow{3}{4em}{Volume L4-S2 [\si{mm^3}]}& \Cord{},  	$V_\mathrm{SpC}$ & $ 25$ (36\%) 	& 4761 (35\%)\\
									& \Fluid{},		 	$V_\mathrm{SpF}$ & $ 31$ (45\%)	& 6323 (47\%)\\
									& \Gb{}, 	 		$V_\mathrm{GB}$ 	& $ 13$ (19\%)	& 2487 (18\%)\\
\hline
\multirow{3}{4em}{Density [\si{g/cm^3}]}& \Cord{}, 	 	$\rho_\mathrm{SpC}$ 	& $ 1.0$ 	& 1.0\\
									& \Fluid{},		 	$\rho_\mathrm{SpF}$ 	& $ 1.0$	& 1.0 \\
									& \Gb{}, 	    	$\rho_\mathrm{GB}$ 	& $ 1.4-1.5$  & 1.0, 1.5, 2.0 \\
\end{tabular}
\end{center}
\vspace{-4mm}
\end{table}

\begin{table}[hptb]
\small
	\caption{\Bpm{} schematic overview. Cross-section views are made at the model centre. Blue areas represent fluid space; contours indicate canal shapes. Triangles with varying background colour/patterns represent \scord{} tissue with varying \gb{} density $\rho_\mathrm{GB}$ from \SIrange{1.0}{2.0}{g/cm^3}. The canal diameter indicates the inner ${\diameter_\mathrm{canal}}$. \Grooves{} are located on the dorsal canal inside; dips are ventral to the \scord{}. Dips and \grooves{} are tested in models with narrow canals only (model 4-7). All models feature a fibre-reinforced \scord{} (short horizontal red line).} 
\label{tab:test_plan}
\begin{center}
\begin{tabularx}{0.85\textwidth}{c|ccccccc}
{Model} number  & 1 & 2 & 3 & 4 & 5 & 6 & 7 \\
  & \pone{\rtwo} & \ptwo{\rtwo} & \pthree{\rtwo} & \pfour{\rtwo} & \pfive{\rtwo} & \psix{\rtwo} & \pseven{\rtwo} \\
\hline
GB density [\si{g/cm^3}] & 1.0 & 1.5 & 2.0 & 1.5 & 1.5 & 1.5 & 1.5 \\
Canal diameter [\si{mm}] & 51 & 51 & 51  & 24 & 24 & 24 & 24 \\
Canal morphology & large & large & large & grooves+dip & dip & grooves & narrow \\
\end{tabularx}
\end{center}
\vspace{-4mm}
\end{table}

\subsection{Locomotion simulation}
\begin{figure*}[thpb]
\centering
\includegraphics[scale=0.6]{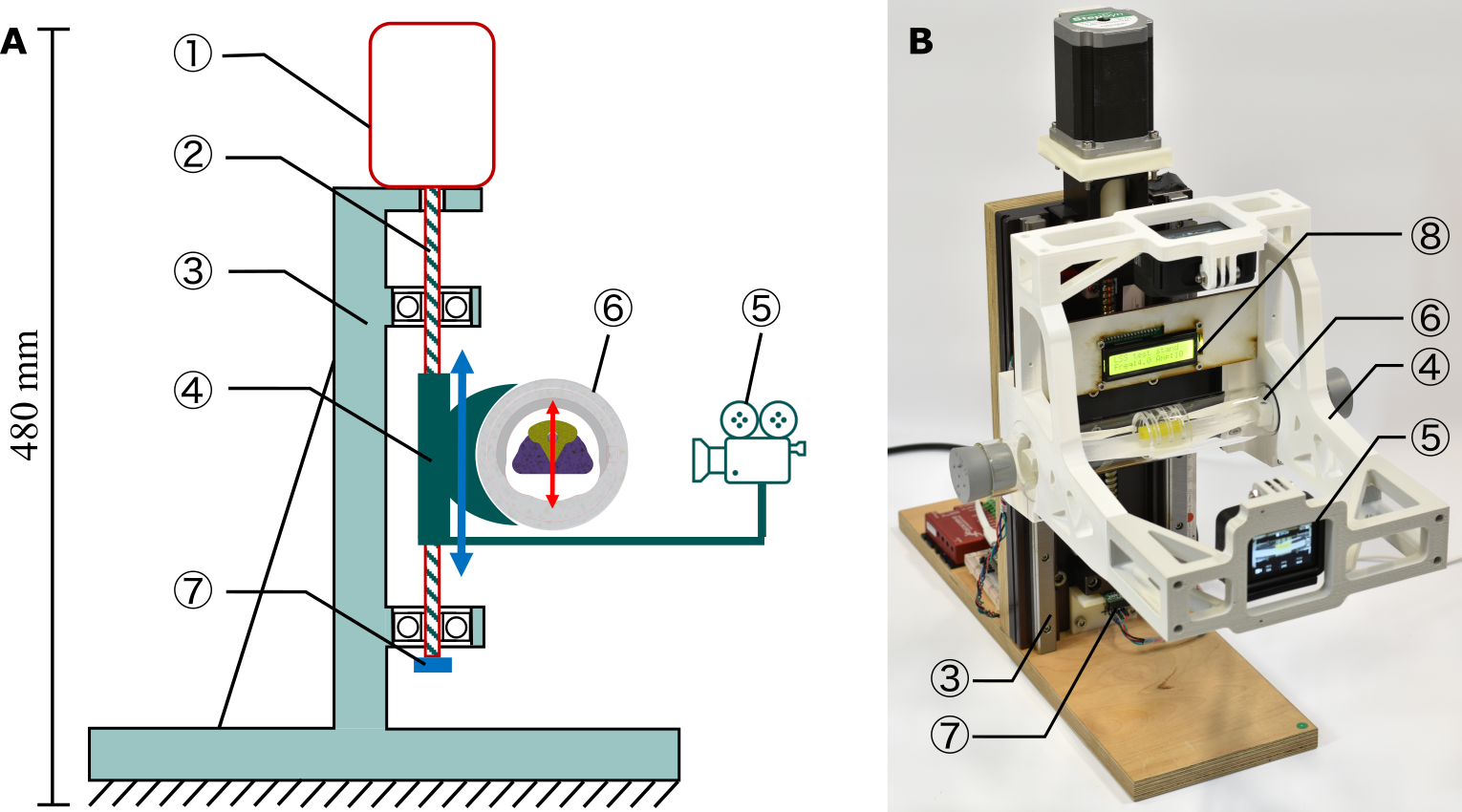}
\caption{The locomotion simulator, schematic (A) and photo (B). The \bpm{}\cnum{6} is mounted to the platform\cnum{4} of a linear drive\cnum{3}. A stepper motor\cnum{1} moves the slider vertically (blue) with a spindle\cnum{2}. An encoder\cnum{7} records the spindle position. A camera\cnum{5} mounted on the moving platform\cnum{4} measures the \bpm{}'s response (red). Video and encoder data are synchronized visually by observing the LCD screen\cnum{8}.
}
\label{fig:oscillator}
\end{figure*}

We developed a locomotion simulator to produce the up-down motion of the bird's trunk during running (\Cref{fig:oscillator}). 
The locomotion simulator generates vertical motions in a controlled manner and records the \bpm{}'s compliant response.
A stepper motor\cnum{1} (103H7823-1740, \textit{Sanyo Denki}) drives a ball screw\cnum{2} (KUHC1205-340-100, \textit{MISUMI}) mounted to a frame made of \SI{20}{mm} plywood\cnum{3}, moving a motion platform\cnum{4} vertically.
The 3D-printed (PLA) motion platform holds the \bpm{}\cnum{6}, an LCD screen\cnum{8} (model 1602) and a video camera\cnum{5} (Hero 5 Black, \textit{GoPro}). 
The camera and the \bpm{} move together. 
Hence, the camera observes the model's compliant response within the local coordinate system. 
A rotary encoder\cnum{7} (AS5045, \textit{AMS}) counts \spindle{} rotations, and the slider displacement is the product of spindle rotation and pitch.

The locomotion simulated is instrumented.
A motor driver (G201X, \textit{geckodrive}) drives the stepper motor. 
A microcontroller (Teensy \num{4.0}, \textit{PJRC}) controls the stepper motor driver and an LCD screen. 
The LCD shows the setup's status.
Encoder data was sampled by a single board computer (Raspberry Pi, v.\,4B), with \SI{10}{\micro\metre} resolution at \SI{1}{kHz} update frequency. 
\Bpm{} movement was camera-recorded at a sampling frequency of \SI{240}{Hz}. 
Both data lines were synchronized by a programmed LCD backlight flash.
The \als{} produces oscillations up to a maximum frequency of \SI{4.5}{Hz} at an amplitude of $\pm \SI{5}{mm}$. 

We found only a few off-the-shelf motion simulators capable of highly dynamic motion (\SIrange{3}{5}{Hz}), all of which were expensive. With this project, we are open-sourcing\footnote{\url{www.github.com/moanan/1_dof_motion_simulator}} our locomotion simulator design and control for barrier-free research; which is capable, easy to replicate, and comparably low-cost.

Testing protocols were identical for all models.
Glass tubes were mounted to the locomotion simulator and vertically driven to oscillate with an amplitude of $\pm \SI{5}{mm}$.
We stopped the motor after \SI{5}{s}.
The resulting damped model motion was recorded for another \SI{3}{s}.
Each model was driven at four `drive' frequencies: \SIlist[list-units = single]{3.0; 3.5; 4.0; 4.5}{Hz}. 
Trials were repeated eight times, resulting in a total of \num{224} trials; 4 \frequencies{} with 8 repetitions and 7 models.

\begin{figure}[hptb]
\centering
\includegraphics[scale=0.9]{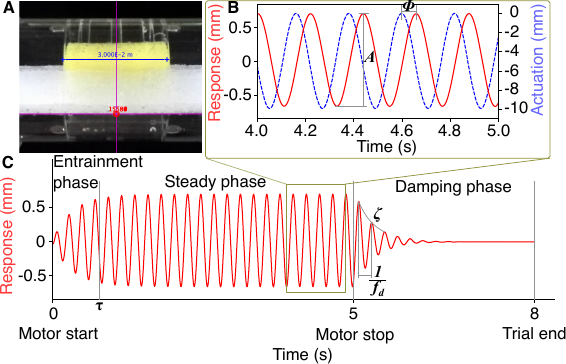}
\caption{Typical model response to external motion. The data shown is extracted from trial 1, model-2, at \SI{4.5}{Hz} external oscillation frequency. Each experiment shows three phases;
\begin{inparaenum}[(1)]
\protect \item The {entrainment phase}, with a settling time $\tau$ spanning from start until the model reaches \SI{90}{\%} of the steady state amplitude. 
\protect \item In the {steady phase} we measure the model's \ppa{} $A$, and the phase shift $\phi$ between the external actuation and the model's response.
\protect \item The {damping phase} starts when the motor is switched off (\SI{5}{s}). We calculate the model's decay rate $\zeta$ and the \dof{} $f_\mathrm{d}$.
\end{inparaenum}
}
\label{fig:time_response_example}
\end{figure}

We extracted the \bpm{}'s movement from the recorded videos with the Tracker software~\citep{noauthor_tracker_nodate}.
An example trial is shown in \Cref{fig:time_response_example}.
We divide the experiment's time series into three phases:\\
\begin{inparaenum}
\phantom{x} \item \PA{} with \st{} $\tau$;\\
\phantom{x} \item \PB{} with \ppa{} $A$, phase shift $\phi$;\\
\phantom{x} \item \PC{} with \rate{} $\zeta$, \dof{} $f_{d}$.\\
\end{inparaenum}
The definitions of the measured parameters in each phase are documented in the supplementary section S2.

\subsection{Functional parameter hypotheses}\label{sec:test_plan}
We tested three hypotheses with seven \bpm{}s (\Cref{tab:test_hypothsis}):\\

\begin{table}[!h]
  \small
  \caption{Hypotheses and the corresponding control variables for model-1 to 7.}
  \label{tab:test_hypothsis}
  \begin{center}
  \begin{tabular}{ccl}
  Hypothesis & Model $\#$ & Controlled variable \\
  \hline
  1  & 1 - 3 	& Density \\
  2  & 2, 7 	& Canal size \\
  3  & 4 - 7 	& Canal morphology \\
  \end{tabular}
  \end{center}
  \vspace{-4mm}
  \end{table}

\begin{inparaenum}[1)]
  \item \textbf{The \gb{} tunes the LSO measurement range}.
  The \gb{} is unique in birds and unexplained.
  Since its density is notably higher than the surrounding \fluid{} and the \scord{}, we expect the \gb{} presents an effectively larger mass leading to higher soft tissue oscillation caused by external movements, compared to a neutrally buoyant \gb{}.
  We compare amplitude response and settling time of three \gb{} densities: \num{1.0}, \num{1.5}, and \SI{2.0}{g/cm^3}, and expect high \gb{} density associates with high amplitude.

  \item \textbf{The narrow \scanal{} dampens soft tissue oscillation}.
  The fluid space that allows for \scord{} oscillation is relatively small.
  Flow resistance increases in the proximity of walls according to the Hagen–Poiseuille equation. 
  Hence, we expect that a narrow \scanal{} increases flow resistance compared to a wide one, leading to reduced soft tissue oscillations.  
  We investigate the effect of large and narrow canal size on the model’s response amplitude and \rate{}.

  \item \textbf{The \scanal{} fine structure diversifies the LSO response}.  
  Previous observations~\citep{kamska20203d,kamska_imaging_nodate,stanchak_balance_2020} hint the distinct \scanal{} morphologies of different birds may be associated with habitats and locomotion modalities.
  We expect the dorsal grooves and the ventral dip~\citep{kamska20203d} both have an effect on spinal fluid flow and soft tissue oscillations.  
  To test this, we map the combinations of the dorsal grooves and the ventral dip, and study the models' response amplitude and \rate{}.
\end{inparaenum}

\section{Results}

\Cref{tab:results} shows all results obtained, ordered by model and \frequency{}, symbols are identical to \Cref{tab:test_plan}. 

\begin{table*}[hptb]
\small
\caption{Results for all models depending on \frequency{}, showing \st{} during \pA{}, \ppa{}, phase shift between drive signal and model oscillation,12 \dof{} and \rate{} during \pC{}. Values are mean values $\pm$ standard error (SE). SEs are not shown if smaller than the rounding digit.}
\label{tab:results}
\begin{center}
\begin{tabularx}{0.7\textwidth}{Xc|lccll} 
Model & Freq. & Settling time	& Amp. & Phase shift & Damped freq. & Decay rate \\
				 & [Hz]			&	\multicolumn{1}{c}{$\tau$ [s]}	&	$A$ [mm]	&	$\phi$ [deg]	&	\multicolumn{1}{c}{$f_{d}$ [Hz]}	&	\multicolumn{1}{c}{$\zeta$ [/]}	\\
\hline
\multirow{4}{*}{1\pone{\rtwo}}  & 3.0 & 0.55 $\pm$ 0.07 & 0.11 & 25 $\pm$ 1   & 4.8           & 0.77 $\pm$ 0.03 \\
                                & 3.5 & 0.39            & 0.25 & 28 $\pm$ 1   & 4.8           & 0.69 $\pm$ 0.02 \\
                                & 4.0 & 0.37 $\pm$ 0.02 & 0.62 & 49 $\pm$ 1   & 4.7           & 0.79            \\
                                & 4.5 & 0.67            & 1.06 & 106 $\pm$ 1  & 4.7           & 0.83            \\ 
\hline
\multirow{4}{*}{2\ptwo{\rtwo}}  & 3.0 & 0.63 $\pm$ 0.05 & 0.13 & 20 $\pm$ 1   & 4.8           & 0.52 $\pm$ 0.02 \\
                                & 3.5 & 0.38            & 0.28 & 25 $\pm$ 1   & 4.8           & 0.59 $\pm$ 0.01 \\
                                & 4.0 & 0.36            & 0.66 & 40 $\pm$ 1   & 4.7           & 0.67 $\pm$ 0.01 \\
                                & 4.5 & 0.78 $\pm$ 0.01 & 1.33 & 92 $\pm$ 1   & 4.7           & 0.73            \\
\hline
\multirow{4}{*}{3\pthree{\rtwo}}& 3.0 & 0.47 $\pm$ 0.04 & 0.19 & 21 $\pm$ 1   & 4.5           & 0.58 $\pm$ 0.02 \\
                                & 3.5 & 0.39            & 0.44 & 24 $\pm$ 1   & 4.4           & 0.66 $\pm$ 0.01 \\
                                & 4.0 & 0.61            & 1.03 & 62 $\pm$ 1   & 4.4           & 0.72            \\
                                & 4.5 & 0.56            & 1.32 & 120 $\pm$ 1  & 4.4           & 0.74 $\pm$ 0.01 \\
\hline
\multirow{4}{*}{4\pfour{\rtwo}} & 3.0 & 0.48            & 0.25 & 46 $\pm$ 1   & 3.7           & 0.82 $\pm$ 0.02 \\
                                & 3.5 & 0.72 $\pm$ 0.02 & 0.57 & 80 $\pm$ 1   & 3.6           & 0.87 $\pm$ 0.01 \\
                                & 4.0 & 0.52            & 0.66 & 130 $\pm$ 1  & 3.6           & 0.87 $\pm$ 0.01 \\
                                & 4.5 & 0.36            & 0.56 & 154 $\pm$ 1  & 3.6           & 0.87 $\pm$ 0.01 \\ 
\hline
\multirow{4}{*}{5\pfive{\rtwo}} & 3.0 & 0.78 $\pm$ 0.16 & 0.06 & 152 $\pm$ 1  & 4.3 $\pm$ 0.3 & 1.28 $\pm$ 0.15 \\
                                & 3.5 & 0.45            & 0.11 & 123 $\pm$ 1  & 4.5 $\pm$ 0.1 & 1.53 $\pm$ 0.16 \\
                                & 4.0 & 0.40            & 0.19 & 127 $\pm$ 1  & 4.3 $\pm$ 0.2 & 1.50 $\pm$ 0.09  \\
                                & 4.5 & 0.42 $\pm$ 0.02 & 0.28 & 147 $\pm$ 1  & 4.4 $\pm$ 0.1 & 1.42 $\pm$ 0.14 \\ 
\hline
\multirow{4}{*}{6\psix{\rtwo}}  & 3.0 & 0.48            & 0.25 & 56 $\pm$ 1   & 3.5           & 0.88 $\pm$ 0.01 \\
                                & 3.5 & 0.70 $\pm$ 0.02 & 0.5  & 103 $\pm$ 1  & 3.5           & 0.92 $\pm$ 0.01 \\
                                & 4.0 & 0.40            & 0.5  & 138 $\pm$ 1  & 3.5           & 0.93 $\pm$ 0.02 \\
                                & 4.5 & 0.37            & 0.48 & 155 $\pm$ 1  & 3.5           & 0.89 $\pm$ 0.02 \\
\hline
\multirow{4}{*}{7\pseven{\rtwo}}& 3.0 & 0.88 $\pm$ 0.22 & 0.07 & 152 $\pm$ 1  & 4.4 $\pm$ 0.2 & 0.88 $\pm$ 0.11 \\
                                & 3.5 & 0.48 $\pm$ 0.04 & 0.12 & 127 $\pm$ 1  & 5.0 $\pm$ 0.2 & 1.24 $\pm$ 0.18 \\
                                & 4.0 & 0.42 $\pm$ 0.02 & 0.19 & 137 $\pm$ 1  & 4.6 $\pm$ 0.2 & 1.25 $\pm$ 0.21 \\
                                & 4.5 & 0.37            & 0.26 & 154 $\pm$ 1  & 4.5 $\pm$ 0.2 & 1.02 $\pm$ 0.13 \\  
\end{tabularx}
\end{center}
\end{table*}

\subsection{Entrainment} 
All models showed an entrainment response similar to \Cref{fig:time_response_example}.
Within \SI{1}{s} models transitioned into steady-state oscillations.
We observed steady-state \pp{} up to \SI{1.3}{mm}.

We observed a distinct frequency peak for each model processed with spectrum analysis. 
The \dof{} measured from \pC{} was independent of the \frequency{} for the same model, but differed between models (\Cref{tab:results}).
\Dofs{} ranged from \SIrange{3.5}{5.0}{Hz}.
Albeit a different calculation method, \dofs{} of model \num{5} and \num{7} showed a similar dependency. 

\subsection{Influence of \gb{} density}
\begin{figure}[!th]
\centering
\includegraphics[scale=1]{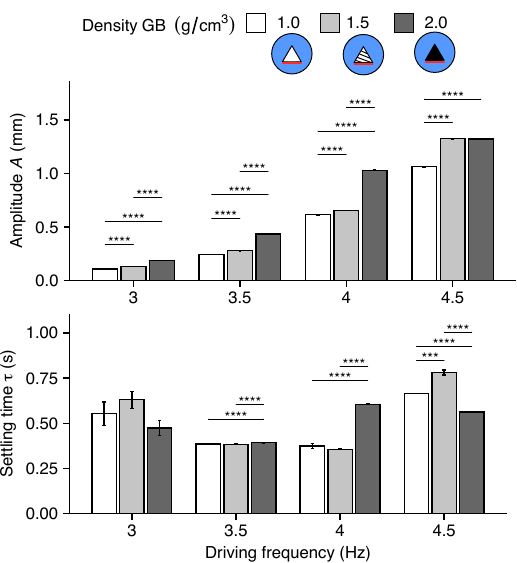}
\caption{Influence of \gb{} density and \frequency{} to \pp{} (A) and settling time (B). Values shown are for model-1, 2, and 3 with density of \num{1.0}, \num{1.5}, and \SI{2.0}{g/cm^3}, respectively. \Frequencies{} tested were \num{3.0}, \num{3.5}, \num{4.0} and \SI{4.5}{Hz}.  Here and in all following figures, standard error shown as error bar, significant \textit{p}-values from \textit{t}-test pairwise comparisons as `****', 0; `***', 0-0.001; `**', 0.001-0.01, `*', 0.1-0.05. Error bars are not shown if the standard error is smaller than the measurement resolution.}
\label{fig:density}
\end{figure}

In general, a denser \gb{} produced a larger \pp{} in the \pB{}. The \pp{} increased with increasing \frequency{} (\Cref{fig:density}A).
The \pp{} significantly differed between \gb{} densities in all models at all \frequencies{} (\textit{t}-test; \textit{p} values \num{<1e-4}) except for densities of \SIrange{1.5}{2.0}{g/cm^3} at \SI{4.5}{Hz}. Model-2 with a \gb{} density of \SI{1.5}{g/cm^3} showed the highest \pp{} of \SI{1.33}{mm}. Independent of \gb{} density, models took between \SIrange{0.36}{0.78}{s} to entrain, and we found no clear tendency for settling time for different \gb{} densities (\Cref{fig:density}B).

\subsection{Influence of canal size} 
\begin{figure}[!th]
  \centering
  \includegraphics[scale=1]{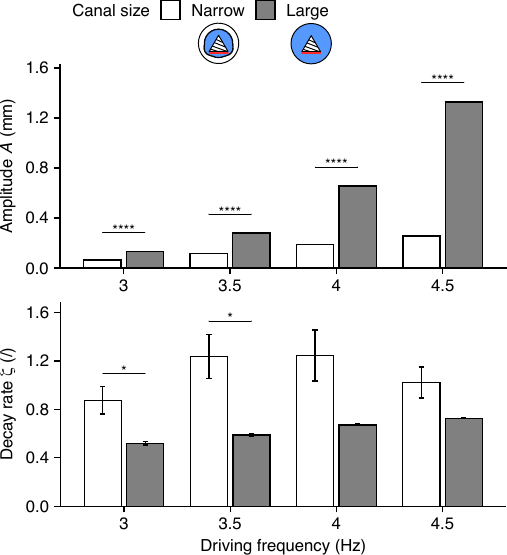}
  \caption{Influence of \scanal{} size and \frequency{} on mean values of \pp{} (A) and decay rate (B). Data is shown for eight trials at \num{3.0}, \num{3.5}, \num{4.0}, and \SI{4.5}{Hz} for model-7 with its narrow-size and model-2 with its large-size canal.
  }
  \label{fig:size}
\end{figure}

The canal size had an effect on the soft tissue response amplitude and its \rate{} (\Cref{fig:size}). 
The larger canal of model-2 yielded a significant higher \pp{} of \SI{1.33}{mm} at \SI{4.5}{Hz}. 
In comparison, the narrow canal of model-7 yielded a maximum amplitude of \SI{0.26}{mm} at \SI{4.5}{Hz} (\textit{t}-test; \textit{p} values \num{<1e-4}). 
For both canal sizes, amplitudes increased with increasing \frequency{}. 
A narrow canal produced damped oscillations with higher \rate{} $\zeta$ of \numrange{0.88}{1.25}, compared to a large canal with \rate{}s between \numrange{0.52}{0.73}. 
However, the difference was only significant at \SI{3.0}{Hz} and \SI{3.5}{Hz} (\textit{t}-test; \textit{p} values \num{<.05}). 
The narrow canal \rate{} did not increase monotonically with \frequency{} as the large canal model did.

\subsection{Influence of canal morphology} 
\begin{figure}[!th]
  \centering
  \includegraphics[scale=1]{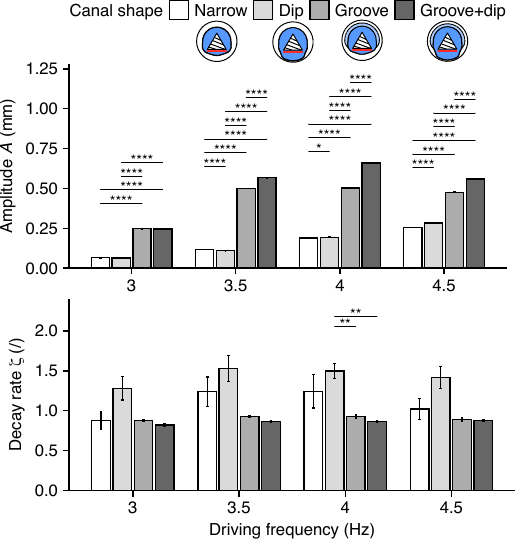}
  \caption{Influence of canal morphology and \frequency{} on mean values of \pp (A) and decay rate (B). Data is shown for eight trials at \num{3.0}, \num{3.5}, \num{4.0}, and \SI{4.5}{Hz} for model-7 with a narrow canal, model-5 with a ventral dip, model-6 with a dorsal groove and model-4 with groove+dip.}
  \label{fig:morphology}
\end{figure}

Varying canal morphologies were implemented by simulating dorsal grooves or a ventral dip (model-4 to 7, \Cref{fig:morphology}). 
The \pp{} significantly differed between these models at all \frequencies{} (\textit{t}-test; \textit{p} values \num{< 0.05}) and it was larger in the presence of grooves+dip with values ranging from \SIrange{0.25}{0.66}{mm}, followed by dorsal grooves with values from \SIrange{0.25}{0.5}{mm}. 
\Rate{}s were lower, between \numrange{0.82}{0.93}, for models with grooves and grooves+dip. 
\Rate{}s were highest between \numrange{1.28}{1.53} for model-5 with its ventral dip. 
\Rate{}s differed significantly only when comparing grooves+dip versus ventral dip and ventral dip versus grooves at \SI{4.0}{Hz} (\textit{t}-test; \textit{p} values \num{<0.05}). 
Note that the \rate{} for the narrow canal and ventral dip models showed comparatively large standard errors due to low \pp{}, with lower signal quality.

\section{Discussion}
We developed a reconfigurable biophysical model of the lumbosacral organ to investigate its physical response during simulated locomotion.
The biophysical model includes the spinal soft tissue, the surrounding \fluid{}, and the \canal{}, with a focus on replicating the key morphologies and material properties relevant to potential mechanosensing functions.
The goal of this work is to produce quantitative results to understand the mechanics of the LSO, especially the influence of the \gb{} and the \canal{}.
We observed typical mass-spring-damper behavior of the soft tissue oscillation, which supports the strain-based accelerometer hypothesis of the LSO~\citep{kamska20203d}.

The function of the LSO remains a debate in the field.
The \gb{} within an enlarged \canal{} is suspicious, and can be only found in birds.
Early studies assumed a ``locomotor brain'' function, due to the large accumulation of nerve cells nearby~\citep{streeter1904structure}.
The \gb{} was further hypothesized to have a nutritional or secretory function~\citep{terni1924ricerche,watterson_development_1949,de1961carbohydrate,azcoitia1985avian}, or relate to myelin synthesis~\citep{benzo_glycogen_1981,de_gennaro_ultrastructural_1976}.
However, these early studies failed to provide a holistic understanding of the LSO, since the specialized morphologies, such as the \canal{} and the \dln{}, were ignored.

Recently, new evidence has supported the mechanosensing function hypothesis of the LSO (\Cref{fig:overview}).
The discovery of mechanoreceptive neurons within the \lobes{} laid the foundation for the potential mechanosensing function~\citep{eide_axonal_1996,eide_development_1996,necker2006specializations,schroeder_specializations_1987,yamanaka_analysis_2012,yamanaka_glutamate_2013,stanchak_molecular_2022}.
The enlarged \canal{} at the LSO region allows for the spinal soft tissue motion~\citep{kamska20203d}, which is essential for stimulating the mechanoreceptors.
During locomotion, the spinal soft tissue is entrained by external acceleration, which forces the \fluid{} to circulate and the \dln{} to strain.
Necker hypothesized that the \flow{}, guided by the \canal{} morphology, may stimulate the mechanoreceptors for a balance function~\citep{necker2006specializations,stanchak_balance_2020,necker2007head,necker_behavioral_2000}.
Instead of the \flow{}, Schroeder and Murray proposed that the ligament strain will transfer to the attached \lobes{} and therefore stimulate the mechanoreceptors~\citep{schroeder_specializations_1987}.
Biological mechanoreceptors are well suited to detect the smallest strain values; they are sensitive in the \si{angstrom} range on a cellular level~\citep{hudspeth_how_1989,barth_mechanics_2019}.
Based on our own observations~\citep{kamska20203d}, the ligament strain can be up to \SI{7.9}{\%}, which is large enough to produce consistent signals.
Although none of these hypotheses have provided conclusive evidence, there were separate studies supported the LSO intraspinal sensing function.
Intraspinal sensing has been identified in lamprey~\citep{mcclellan1993mechanosensory,grillner_edge_1984} and zebrafish~\citep{bohm_csfcontacting_2016,picton_spinal_2021}, where their mechanosensors' arrangement is similar to birds' \lobes{}.
Another study suggested that balance sensing exists in the bird's body.
After labyrinthectomy and spinal cord transection, birds can still reflexively compensate for body rotations without the vision and vestibular sensing~\citep{Thorson1973a}.
Here, the mechanosensing function of the LSO is a potential explanation.

The hypothesized LSO sensing mechanism can be essential for birds' exceptional locomotion agility.
As agile locomotion requires fast sensing and action, the LSO potentially provides fast state feedback by minimizing the time required for detecting accelerations and sending the corresponding signals (\Cref{fig:overview}).
On one hand, the body accelerations, as a result of contact with environments, can be transmitted to the \canal{} and the LSO through bones in the form of shock wave.
The shock wave transmitted through bones~\citep{pelker1983stress,miller2020control} is at least one order of magnitudes faster than the nerve conduction speed~\citep{more2010scaling}, allowing for fast acceleration detection.
On the other hand, the close integration of LSO to the \cord{} greatly reduces the nerve conduction delays, and the output signals are likely integrated to the central pattern generator circuits for limb coordination~\citep{eide_axonal_1996,eide_development_1996,stanchak_molecular_2022}.
Moreover, the fast state feedback is increasingly important for larger birds.
An extreme example is the ancestor of birds---dinosaur, as the extended nerve fiber may prohibit in-time muscle response.
Coincidently, recent fossil records of dinosaurs have identified a lumbosacral canal enlargement similar to the birds'~\citep{wedel2021expanded}, suggesting the importance of the LSO for evolutionary success.

While the mechanosensing hypothesis has been well established~\citep{necker2006specializations,schroeder_specializations_1987,kamska20203d,stanchak_balance_2020}, several key processes remain unsolved.
First, whether the spinal soft tissue can move relative to the \canal{} is still questionable.
So far, we have been able to observe only a small amount of soft tissue motion in chicken~\citep{kamska_imaging_nodate}.
Second, assuming the soft tissue can move, can truck movement entrain the spinal soft tissue motion? Further, how the morphologies and material properties shape the LSO response?
In this work, we aim to understand this entrainment behavior.
Third, supposing the soft tissue entrainment exists, how the neural circuits process the mechanoreceptors' signals and how the signals can be mapped to what kinds of sensing modalities is still unknown.

To answer the above questions, the main challenge is the inaccessibility of the LSO.
Birds' \canal{} is densely fused, making in vivo measurement of the soft tissue movement and the mechanosensor activity almost impossible.
Numerical simulation is not viable due to the complex interaction among the viscoelastic tissue, the \fluid{}, and the rigid \canal{}.
Moreover, large deformation of viscoelastic materials are poorly predicted in simulation.
Alternatively, we propose using biophysical simulation to study the entrainment mechanics of the LSO.
Biophysical models are effective tools for testing the underlying mechanisms of biological systems~\citep{panjabi1998cervical,poel2020assessing}.
Benefiting from soft robotic techniques, our LSO biophysical model is parametric, modular, and based on precise anatomical data~\citep{kamska20203d} (\Cref{fig:phantom}).
Testing different configurations of the biophysical model on our custom-developed locomotion simulator (\Cref{fig:oscillator}) revealed how individual morphology and material property affect the LSO response (\Cref{tab:results}).

The biophysical model confirmed the mobility of the soft tissue.
Although the model is not a one-to-one copy of its biological reference, the underlining physics is the same.
The models were carefully crafted to account for the appropriate physical forces acting on the model, such as the gravitational forces, buoyant forces, locomotion accelerations.
As long as we are able to measure notable soft tissue motion, we can trust a similar soft tissue motion will exist in the biological LSO.
As expected, all models have shown typical mass-spring-damper response (\Cref{fig:time_response_example}).
In \pA{}, the \st{} conveys how fast a model responds to external oscillation and provides an intuition of the model's temporal sensitivity.
In \pB{}, assuming the soft tissue displacement is in proportional to the mechanosensing signal intensity, the \ppa{} indicates the signal strength.
In \pC{}, the \rate{} shows how fast the LSO resets after a locomotion stop.
We consider that an effective mechanosensor rapidly fades out oscillations through fluid damping and lossy tissue deformations.
Here, we observed a distinct \dof{}, ranged between \SIrange{3.5}{4.5}{Hz}, which overlaps with locomotion frequencies of running birds of \SIrange{3}{5}{Hz}~\citep{gatesy1991bipedal,Smith_2010,Daley2018scaling}.
Note that the \dof{} is an approximation of the resonant frequency \citep{morin2021}.
A resonating LSO will increase the oscillation amplitude, potentially increasing the sensor signal intensity.
We did not perform a system identification test to determine the precise mass-spring-damper parameters, as our goal was to understand the impact of morphological variations to the LSO response, rather than systematically investigate an artificial system.
In particular, we focus on the \gb{} density, the \canal{} size and morphology.

The \gb{} density showed notable influence on the soft tissue oscillation (\Cref{fig:density}).
We implemented the \gb{} denser than the \cord{} and the \fluid{}. 
In simplified mass-spring-damper systems, higher masses tend to oscillate at lower frequencies. 
We indeed observed a small reduction in the \dof{} between model-2 and 3 with higher \gb{} densities; from averaged \SI{4.8}{Hz} to \SI{4.4}{Hz}, which again confirmed the mass-spring-damper behavior.
The increased \gb{} density is associated with higher \ppa{} (\Cref{fig:density}A).
Since the \ppa{} is assumed to connect with signal strength, the denser \gb{} could act as a signal amplifier.
Consequently, running birds that experience higher vertical accelerations could feature a low-density \gb{}, leading to a signal intensity similar to flying or swimming birds with low-acceleration locomotion modes.
As such, the \gb{} density could adjust the acceleration measurement range, i.e., amplifying small acceleration with high density, or suppressing excessive acceleration with low density.
Varying \gb{} densities could have developed as a trade-off between LSO responsiveness and protection.
Sudden perturbations during locomotion can lead to high accelerations, potentially damaging the \cord{} tissue. 
\Fluid{} and \ligaments{} would reduce soft tissue motions through fluid buoyancy, damping, and ligament elasticity~\citep{telano_physiology_2022}.
We think buoyancy could protect the best if tissue densities are similar to the surrounding fluid.
Too high or low density would lead to sinking or flotation. 
Hence, two or more opposing motivations exist; feature a neutrally buoyant density to protect the \scord{}, and increase \gb{} density for sufficiently high \ppa{}.

The \canal{} dimensions with submillimeter flow classifies as a micro-fluid environment, with high fluid resistance slowing the flow and dampening oscillations.
The dimension of lumbosacral canal expansion has been a central question in the functional hypothesis of the LSO~\citep{emmert1811beobachtungen,necker_structure_2005,necker_specializations_1999,kamska20203d,stanchak_balance_2020}.
As expected, a narrow \canal{} led to smaller \ppa{} and higher \rate{} as a result of high damping, in contrast to a large-diameter \canal{} (\Cref{fig:size}).
The high damping has two main effects on the LSO sensing mechanism.
First, it suppresses excessive \cord{} deflection during high acceleration events, such as strong perturbations, protecting the \cord{} tissue. 
Second, the resulting high \rate{} enhances fast sensor reset.
When locomotion stops, the damped oscillation will continue simulating the mechanosensors and generating signals.
In this scenario, these signals may become noise and should be minimized as soon as possible.
Therefore, the \canal{} expansion could have been optimized for the damping term of a mass-spring-damper system.

Additionally, the \canal{} enlargement features different fine structure among different birds~\citep{kamska20203d,stanchak_balance_2020,kamska_imaging_nodate}.
From our preliminary scanning data of swan, pigeon, quail and chicken, selected to represent diverse locomotion modalities, we observed different shape of \grooves{} and \dip{}.
We studied whether these fine structures would play an important role in the LSO response by testing the combinations of \grooves{} and \dip{} (\Cref{fig:morphology}).
Vertical \cord{} motions will displace fluid inside the volume-constant \canal{}; when the \cord{} moves up, the \fluid{} is pushed down.
We can assume that fluid of the same volume is moved against the \cord{} motion.
Lateral gaps and, possibly, \grooves{} will guide the flow between the \cord{} and the inner \canal{}~\citep{necker2006specializations,stanchak_balance_2020}.
In quails~\citep{kamska20203d}, only small flow space is available laterally---between \SIrange{0.3}{0.8}{mm}---which we mimicked in model-4 to 7 (\Cref{tab:test_plan}).
Models with extra dorsal structures showed larger \ppa{} and lower \rate{}s, compared to canals without~(\Cref{fig:morphology}). 
Hence, the laterally and dorsally extending \grooves{} could act as fluid reliefs or flow channels~\citep{kamska20203d,stanchak_balance_2020,necker_behavioral_2000}.
Model-5 featuring the \dip{} behaved somewhat unexpectedly. 
Albeit a larger fluid space, it caused a higher \rate{} compared to the narrow-canal model-7 (\Cref{fig:morphology}).
We conclude that model-5's \dip{} might have dampened the oscillations.
Although the \canal{} we developed is highly simplified compared to our reference quail model (\Cref{fig:phantom}), small modifications to the canal morphology has already led to considerable different responses (\Cref{fig:morphology}).
Hence, the \canal{} fine structure diversifies the LSO response, likely associated with the locomotion modality of birds.
For example, we expect more pronounced \grooves{} and \dip{} for swimmers and divers, since the lack of visual cues and low body acceleration might require higher sensitivity.
To better understand the function of \canal{} fine structure in relation to locomotion modality and LSO response, a more realistic \canal{} modelling is required.

While our three hypotheses have been validated, there are several directions for future improvement.
Our locomotion simulator (\Cref{fig:oscillator}) is limited to only one degree of freedom (DoF), while real-world locomotion acceleration is in 3D space and has six DoFs.
This was because suitable motion simulators capable of highly dynamic motion in six DoFs were not affordable.
As a starting point, we custom-developed our own motion simulator, and open sourced the design for barrier-free research.
Nevertheless, this platform was sufficient to prove the feasibility of our biophysical simulation framework.
In the future, a 6-DoF motion simulator~\citep{pradhan2023upside} will allow testing the LSO response under rolling, yawing, pitching, etc.
We also expect to improve the \bpm{} design by instrumentation.
By adding strain or pressure sensors at the location of the mechanoreceptors (\Cref{fig:overview}), testing the differential mechanosensing mechanism~{\citep{kamska20203d}} will be possible.
If we can successfully map the strain or pressure signals to the body accelerations in six DoFs, we close the loop of the mechanosensing hypothesis as shown in \Cref{fig:overview}.
More importantly, the improved and instrumented framework will permit the correlation of LSO sensitivity on specific DoF.
For instance, the topology of the mechanosensors in LSO may have been optimized to predict heaving, pitching, and rolling, which are the dominant motions for most birds.

Overall, our biophysical simulation framework emphasizes the notion of understanding through creation, also known as ``What I cannot create, I do not understand.''
We expect creating the ``physical twins'' of the LSO will be a power tool to improve our understanding of it.

\section{Conclusions}
We developed a simplified, modular, biophysical model of the lumbosacral organ of birds to advance our understanding of this system.
Models were mounted to an instrumented setup that simulated vertical oscillations and recorded the model response.
We presented results that support the central hypothesis that external motion, such as running, leads to measurable LSO \cord{} movement.
The \gb{} density has a strong impact on the LSO response intensity.
We found that narrow \canal{} reduces soft tissue motions through the effects of damping, effectively protecting the \scord{}.
The \canal{} fine structure potentially associated with locomotion modalities of different birds.
In this work, we focus on understanding the mechanics of the LSO.
In the future, we expect to test the mechanosensing mechanism with a more elaborate LSO model and a 6-DoF locomotion simulator.

\textbf{Acknowledgements:}
This work was supported by the China Scholarship Council (CSC) and the International Max Planck Research School for Intelligent Systems (IMPRS-IS).

\textbf{Data availability:}
The biophysical model design has been uploaded as part of the supplementary material.

\printbibliography

@article{azcoitia1985avian,
    author = {Azcoitia, I and Fernandez-Soriano, J and Fernandez-Ruiz, B},
    journal = {Journal fur Hirnforschung},
    number = {6},
    pages = {651--657},
    title = {Is the avian glycogen body a secretory organ?},
    volume = {26},
    year = {1985}
}

@article{barth_mechanics_2019,
    author = {Barth, Friedrich G.},
    doi = {10.1007/s00359-019-01355-z},
    issn = {0340-7594},
    journal = {Journal of Comparative Physiology. A, Neuroethology, Sensory, Neural, and Behavioral Physiology},
    number = {5},
    pages = {661--686},
    pmcid = {PMC6726712},
    pmid = {31270587},
    title = {Mechanics to pre-process information for the fine tuning of mechanoreceptors},
    url = {https://www.ncbi.nlm.nih.gov/pmc/articles/PMC6726712/},
    volume = {205},
    year = {2019}
}

@techreport{bausch_spezialisierungen_2014,
    author = {Bausch, Pia},
    title = {Die {Spezialisierungen} des lumbosakralen {Wirbelkanals} beim {Großen} {Emu} ({Dromaius} novaehollandiae)},
    year = {2014}
}

@article{bekoff_coordinated_1975,
    author = {Bekoff, A. and Stein, P. S. and Hamburger, V.},
    doi = {10.1073/pnas.72.4.1245},
    issn = {0027-8424, 1091-6490},
    journal = {Proceedings of the National Academy of Sciences},
    month = {April},
    number = {4},
    pages = {1245--1248},
    title = {Coordinated motor output in the hindlimb of the 7-day chick embryo},
    url = {http://www.pnas.org/content/72/4/1245},
    volume = {72},
    year = {1975}
}

@article{benzo_glycogen_1981,
    author = {Benzo, Camillo A. and De Gennaro, Louis D.},
    doi = {10.1002/jez.1402150106},
    issn = {1097-010X},
    journal = {Journal of Experimental Zoology},
    number = {1},
    pages = {47--52},
    shorttitle = {Glycogen metabolism in the developing accessory lobes of {Lachi} in the nerve cord of the chick},
    title = {Glycogen metabolism in the developing accessory lobes of {Lachi} in the nerve cord of the chick: {Metabolic} correlations with the avian glycogen body},
    url = {http://onlinelibrary.wiley.com/doi/10.1002/jez.1402150106/abstract},
    volume = {215},
    year = {1981}
}

@article{benzo_hypothesis_1983,
    author = {Benzo, C. A. and De Gennaro, L. D.},
    doi = {10.1016/0306-9877(83)90053-1},
    issn = {0306-9877},
    journal = {Medical Hypotheses},
    month = {January},
    number = {1},
    pages = {69--76},
    shorttitle = {An hypothesis of function for the avian glycogen body},
    title = {An hypothesis of function for the avian glycogen body: {A} novel role for glycogen in the central nervous system},
    volume = {10},
    year = {1983}
}

@article{berthouze_assembly_2008,
    author = {Berthouze, Luc and Goldfield, Eugene C.},
    doi = {10.1002/icd.542},
    journal = {Infant and Child Development},
    number = {1},
    pages = {25--42},
    title = {Assembly, tuning, and transfer of action systems in infants and robots},
    url = {http://dx.doi.org/10.1002/icd.542},
    volume = {17},
    year = {2008}
}

@article{bohm_csfcontacting_2016,
    author = {Böhm, Urs Lucas and Prendergast, Andrew and Djenoune, Lydia and Nunes Figueiredo, Sophie and Gomez, Johanna and Stokes, Caleb and Kaiser, Sonya and Suster, Maximilliano and Kawakami, Koichi and Charpentier, Marine and Concordet, Jean-Paul and Rio, Jean-Paul and Del Bene, Filippo and Wyart, Claire},
    doi = {10.1038/ncomms10866},
    issn = {2041-1723},
    journal = {Nature Communications},
    month = {March},
    number = {1},
    pages = {10866},
    title = {{CSF}-contacting neurons regulate locomotion by relaying mechanical stimuli to spinal circuits},
    volume = {7},
    year = {2016}
}

@article{conway_proprioceptive_1987,
    author = {Conway, B. A. and Hultborn, H. and Kiehn, O.},
    doi = {10.1007/BF00249807},
    issn = {0014-4819, 1432-1106},
    journal = {Experimental Brain Research},
    month = {November},
    number = {3},
    pages = {643--656},
    title = {Proprioceptive input resets central locomotor rhythm in the spinal cat},
    url = {http://link.springer.com/article/10.1007/BF00249807},
    volume = {68},
    year = {1987}
}

@article{daley2006running,
    author = {Daley, Monica A and Biewener, Andrew A},
    doi = {10.1073/pnas.0601473103},
    journal = {Proceedings of the National Academy of Sciences},
    number = {42},
    pages = {15681--15686},
    publisher = {National Acad Sciences},
    title = {Running over rough terrain reveals limb control for intrinsic stability},
    volume = {103},
    year = {2006}
}

@article{daley2009role,
    author = {Daley, Monica A and Voloshina, Alexandra and Biewener, Andrew A},
    doi = {10.1113/jphysiol.2009.171017},
    journal = {The Journal of physiology},
    number = {11},
    pages = {2693--2707},
    publisher = {Wiley Online Library},
    title = {The role of intrinsic muscle mechanics in the neuromuscular control of stable running in the guinea fowl},
    volume = {587},
    year = {2009}
}

@article{Daley2018scaling,
    author = {Daley, Monica A and Birn-Jeffery, Aleksandra},
    doi = {10.1242/jeb.152538},
    issn = {0022-0949},
    journal = {Journal of Experimental Biology},
    number = {10},
    pages = {jeb152538},
    title = {Scaling of avian bipedal locomotion reveals independent effects of body mass and leg posture on gait},
    volume = {221},
    year = {2018}
}

@article{de1961carbohydrate,
    author = {De Gennaro, Louis D},
    doi = {10.2307/1539536},
    journal = {The Biological Bulletin},
    number = {3},
    pages = {348--352},
    publisher = {Marine Biological Laboratory},
    title = {The carbohydrate composition of the glycogen body of the chick embryo as revealed by paper chromatography},
    volume = {120},
    year = {1961}
}

@article{de_gennaro_ultrastructural_1976,
    author = {De Gennaro, Louis D. and Benzo, Camillo A.},
    doi = {10.1002/jez.1401980111},
    issn = {1097-010X},
    journal = {Journal of Experimental Zoology},
    number = {1},
    pages = {97--107},
    title = {Ultrastructural characterization of the accessory lobes of {Lachi} ({Hofmann}'s nuclei) in the nerve cord of the chick. {I}. {Axoglial} synapses},
    url = {http://onlinelibrary.wiley.com/doi/10.1002/jez.1401980111/abstract},
    volume = {198},
    year = {1976}
}

@article{eide_axonal_1996,
    author = {Eide, Anne Lill},
    doi = {10.1007/BF00187926},
    issn = {0340-2061, 1432-0568},
    journal = {Anatomy and Embryology},
    month = {June},
    number = {6},
    pages = {543--557},
    title = {The axonal projections of the {Hofmann} nuclei in the spinal cord of the late stage chicken embryo},
    url = {http://link.springer.com/article/10.1007/BF00187926},
    volume = {193},
    year = {1996}
}

@article{eide_development_1996,
    author = {Eide, Anne Lill and Glover, Joel C.},
    doi = {10.1523/JNEUROSCI.16-18-05749.1996},
    issn = {0270-6474, 1529-2401},
    journal = {The Journal of Neuroscience},
    month = {September},
    number = {18},
    pages = {5749--5761},
    title = {Development of an {Identified} {Spinal} {Commissural} {Interneuron} {Population} in an {Amniote}: {Neurons} of the {Avian} {Hofmann} {Nuclei}},
    url = {http://www.jneurosci.org/content/16/18/5749},
    volume = {16},
    year = {1996}
}

@article{emmert1811beobachtungen,
    author = {Emmert, AGF},
    journal = {Reil's Arch. Physiol.},
    pages = {377--392},
    title = {Beobachtungen {\"u}ber einige anatomische Eigenheiten der V{\"o}gel},
    volume = {10},
    year = {1811}
}

@article{gatesy1991bipedal,
    author = {Gatesy, SM and Biewener, AA},
    doi = {10.1111/j.1469-7998.1991.tb04794.x},
    journal = {Journal of Zoology},
    number = {1},
    pages = {127-147},
    publisher = {Wiley Online Library},
    title = {Bipedal locomotion: effects of speed, size and limb posture in birds and humans},
    volume = {224},
    year = {1991}
}

@article{goldfield_infant_1993,
    author = {Goldfield, Eugene C. and Kay, Bruce A. and Warren, William H.},
    doi = {10.2307/1131330},
    issn = {00093920},
    journal = {Child Development},
    month = {August},
    number = {4},
    pages = {1128},
    shorttitle = {Infant {Bouncing}},
    title = {Infant {Bouncing}: {The} {Assembly} and {Tuning} of {Action} {Systems}},
    url = {http://www.jstor.org/discover/10.2307/1131330?uid=3737864&uid=2&uid=4&sid=21104976219167},
    volume = {64},
    year = {1993}
}

@article{grillner_edge_1984,
    author = {Grillner, S. and Williams, T. and Lagerback, P. A.},
    doi = {10.1126/science.6691161},
    issn = {0036-8075, 1095-9203},
    journal = {Science},
    month = {February},
    number = {4635},
    pages = {500--503},
    title = {The edge cell, a possible intraspinal mechanoreceptor},
    url = {http://www.sciencemag.org/content/223/4635/500},
    volume = {223},
    year = {1984}
}

@article{haen_whitmer_mixed_2021,
    author = {Haen Whitmer, Karri},
    doi = {10.31274/isudp.2021.67},
    month = {February},
    title = {A {Mixed} {Course}-{Based} {Research} {Approach} to {Human} {Physiology}},
    url = {https://openlibrary-repo.ecampusontario.ca/jspui/handle/123456789/881},
    year = {2021}
}

@article{hudspeth_how_1989,
    author = {Hudspeth, A. James},
    doi = {10.1038/341397a0},
    journal = {Nature},
    note = {Publisher: Nature Publishing Group},
    number = {6241},
    pages = {397--404},
    title = {How the ear's works work},
    volume = {341},
    year = {1989}
}

@article{kamska20203d,
    author = {Kamska, Viktoriia and Daley, Monica and Badri-Spröwitz, Alexander},
    doi = {10.1093/iob/obaa037},
    issn = {2517-4843},
    journal = {Integrative Organismal Biology},
    month = {January},
    number = {obaa037},
    title = {{3D} {Anatomy} of the {Quail} {Lumbosacral} {Spinal} {Canal}—{Implications} for {Putative} {Mechanosensory} {Function}},
    url = {https://doi.org/10.1093/iob/obaa037},
    volume = {2},
    year = {2020}
}

@article{kamska_imaging_nodate,
    author = {Kamska, Viktoriia and Mo, An and Pohmann, Rolf and Karakostis, Fotios Alexandros and Daley, Monica A. and Badri-Spr\"owitz, Alexander},
    notes = {submitting},
    title = {Imaging the soft tissues motion inside the canal, submitting}
}

@article{knuesel_effects_2011,
    author = {Knuesel, Jeremie and Ijspeert, Auke J.},
    doi = {10.1186/1471-2202-12-S1-P158},
    issn = {1471-2202},
    journal = {BMC Neuroscience},
    month = {7},
    number = {Suppl 1},
    pages = {P158},
    title = {Effects of muscle dynamics and proprioceptive feedback on the kinematics and {CPG} activity of salamander stepping},
    url = {http://www.biomedcentral.com/1471-2202/12/S1/P158/},
    volume = {12},
    year = {2011}
}

@incollection{kuchenbecker_verrotouch_2010,
    author = {Kuchenbecker, Katherine J. and Gewirtz, Jamie and McMahan, William and Standish, Dorsey and Martin, Paul and Bohren, Jonathan and Mendoza, Pierre J. and Lee, David I.},
    booktitle = {Haptics: {Generating} and {Perceiving} {Tangible} {Sensations}},
    doi = {10.1007/978-3-642-14064-8_28},
    editor = {Kappers, Astrid M. L. and Erp, Jan B. F. van and Tiest, Wouter M. Bergmann and Helm, Frans C. T. van der},
    isbn = {978-3-642-14063-1},
    pages = {189--196},
    publisher = {Springer Berlin Heidelberg},
    shorttitle = {{VerroTouch}},
    title = {{VerroTouch}: {High}-{Frequency} {Acceleration} {Feedback} for {Telerobotic} {Surgery}},
    url = {http://link.springer.com/chapter/10.1007/978-3-642-14064-8_28},
    year = {2010}
}

@article{mcclellan1993mechanosensory,
    author = {McClellan, ANDREW D and Jang, WOOCHAN},
    doi = {10.1152/jn.1994.71.6.1-a},
    journal = {Journal of Neurophysiology},
    number = {6},
    pages = {2442--2454},
    title = {Mechanosensory inputs to the central pattern generators for locomotion in the lamprey spinal cord: resetting, entrainment, and computer modeling},
    volume = {70},
    year = {1993}
}

@article{miller2020control,
    author = {Miller, Thomas E and Mortimer, Beth},
    doi = {10.3389/fevo.2020.587846},
    journal = {Frontiers in Ecology and Evolution},
    pages = {587846},
    publisher = {Frontiers Media SA},
    title = {Control vs. constraint: understanding the mechanisms of vibration transmission during material-bound information transfer},
    volume = {8},
    year = {2020}
}

@article{moller_blood_brain_2003,
    author = {Möller, Wilhelm and Kummer, Wolfgang},
    doi = {10.1007/s00441-003-0742-0},
    issn = {0302-766X, 1432-0878},
    journal = {Cell and Tissue Research},
    month = {July},
    number = {1},
    pages = {71--80},
    title = {The blood-brain barrier of the chick glycogen body (corpus gelatinosum) and its functional implications},
    url = {http://link.springer.com/article/10.1007/s00441-003-0742-0},
    volume = {313},
    year = {2003}
}

@article{more2010scaling,
    author = {More, Heather L and Hutchinson, John R and Collins, David F and Weber, Douglas J and Aung, Steven KH and Donelan, J Maxwell},
    doi = {10.1098/rspb.2010.0898},
    journal = {Proceedings of the Royal Society B: Biological Sciences},
    number = {1700},
    pages = {3563--3568},
    publisher = {The Royal Society},
    title = {Scaling of sensorimotor control in terrestrial mammals},
    volume = {277},
    year = {2010}
}

@article{more2018scaling,
    author = {More, Heather L and Donelan, J Maxwell},
    doi = {10.1098/rspb.2018.0613},
    journal = {Proceedings of the Royal Society B},
    number = {1885},
    pages = {20180613},
    publisher = {The Royal Society},
    title = {Scaling of sensorimotor delays in terrestrial mammals},
    volume = {285},
    year = {2018}
}

@inbook{morin2021,
    author = {Morin, David},
    booktitle = {Waves},
    chapter = {1.2},
    doi = {10.1017/CBO9780511808951.005},
    pages = {15},
    title = {Oscillations},
    url = {https://scholar.harvard.edu/david-morin/waves},
    year = {2021}
}

@article{mouel_anticipatory_2019,
    author = {Mouel, Charlotte Le and Brette, Romain},
    doi = {10.1371/journal.pcbi.1007463},
    issn = {1553-7358},
    journal = {PLOS Computational Biology},
    month = {November},
    note = {Publisher: Public Library of Science},
    number = {11},
    pages = {e1007463},
    title = {Anticipatory coadaptation of ankle stiffness and sensorimotor gain for standing balance},
    url = {https://journals.plos.org/ploscompbiol/article?id=10.1371/journal.pcbi.1007463},
    volume = {15},
    year = {2019}
}

@article{necker2006specializations,
    author = {Necker, Reinhold},
    doi = {10.1007/s00359-006-0105-x},
    journal = {Journal of Comparative Physiology A},
    number = {5},
    pages = {439},
    publisher = {Springer},
    title = {Specializations in the lumbosacral vertebral canal and spinal cord of birds: evidence of a function as a sense organ which is involved in the control of walking},
    volume = {192},
    year = {2006}
}

@article{necker2007head,
    author = {Necker, Reinhold},
    doi = {10.1007/s00359-007-0281-3},
    journal = {Journal of comparative physiology A},
    number = {12},
    pages = {1177},
    publisher = {Springer},
    title = {Head-bobbing of walking birds},
    volume = {193},
    year = {2007}
}

@article{necker_behavioral_2000,
    author = {Necker, R. and Jan{\ss}en, A. and Beissenhirtz, T.},
    doi = {10.1007/s003590050440},
    issn = {0340-7594, 1432-1351},
    journal = {Journal of Comparative Physiology A},
    month = {4},
    number = {4},
    pages = {409--412},
    title = {Behavioral evidence of the role of lumbosacral anatomical specializations in pigeons in maintaining balance during terrestrial locomotion},
    url = {http://link.springer.com/article/10.1007/s003590050440},
    volume = {186},
    year = {2000}
}

@article{necker_specializations_1999,
    author = {Necker, R.},
    doi = {10.1076/ejom.37.2.211.4758},
    issn = {0924-3860},
    journal = {European Journal of Morphology},
    month = {4},
    number = {2-3},
    pages = {211--214},
    shorttitle = {Specializations in the {Lumbosacral} {Spinal} {Cord} of {Birds}},
    title = {Specializations in the {Lumbosacral} {Spinal} {Cord} of {Birds}: {Morphological} and {Behavioural} {Evidence} for a {Sense} of {Equilibrium}},
    url = {http://europepmc.org/abstract/MED/10342459},
    volume = {37},
    year = {1999}
}

@article{necker_structure_2005,
    author = {Necker, R.},
    doi = {10.1007/s00429-005-0016-6},
    issn = {0340-2061, 1432-0568},
    journal = {Anatomy and Embryology},
    month = {8},
    number = {1},
    pages = {59--74},
    title = {The structure and development of avian lumbosacral specializations of the vertebral canal and the spinal cord with special reference to a possible function as a sense organ of equilibrium},
    url = {http://link.springer.com/article/10.1007/s00429-005-0016-6},
    volume = {210},
    year = {2005}
}

@misc{noauthor_tracker_nodate,
    author = {Tracker Software},
    title = {Tracker {Video} {Analysis} and {Modeling} {Tool} for {Physics} {Education}},
    url = {https://physlets.org/tracker/}
}

@article{panjabi1998cervical,
    author = {Panjabi, Manohar M},
    doi = {10.1097/00007632-199812150-00007},
    journal = {Spine},
    number = {24},
    pages = {2684-2699},
    publisher = {LWW},
    title = {Cervical spine models for biomechanical research},
    volume = {23},
    year = {1998}
}

@article{pelker1983stress,
    author = {Pelker, Richard R and Saha, Subrata},
    doi = {10.1016/0021-9290(83)90062-3},
    journal = {Journal of Biomechanics},
    number = {7},
    pages = {481--489},
    publisher = {Elsevier},
    title = {Stress wave propagation in bone},
    volume = {16},
    year = {1983}
}

@article{picton_spinal_2021,
    author = {Picton, Laurence D. and Bertuzzi, Maria and Pallucchi, Irene and Fontanel, Pierre and Dahlberg, Elin and Björnfors, E. Rebecka and Iacoviello, Francesco and Shearing, Paul R. and El Manira, Abdeljabbar},
    doi = {10.1016/j.neuron.2021.01.018},
    issn = {0896-6273},
    journal = {Neuron},
    month = {April},
    number = {7},
    pages = {1188--1201.e7},
    title = {A spinal organ of proprioception for integrated motor action feedback},
    volume = {109},
    year = {2021}
}

@article{poel2020assessing,
    author = {Poel, R and Belosi, F and Albertini, F and Walser, M and Gisep, A and Lomax, AJ and Weber, DC},
    journal = {Physics in Medicine \& Biology},
    number = {24},
    pages = {245031},
    publisher = {IOP Publishing},
    title = {Assessing the advantages of CFR-PEEK over titanium spinal stabilization implants in proton therapy—a phantom study},
    volume = {65},
    year = {2020}
}

@article{pradhan2023upside,
    author = {Pradhan, Nayan Man Singh and Frank, Patrick and Mo, An and Badri-Spr{\"o}witz, Alexander},
    doi = {10.48550/arXiv.2303.17974},
    journal = {arXiv preprint arXiv:2303.17974},
    title = {Upside down: affordable high-performance motion platform},
    year = {2023}
}

@article{Rosenberg_Necker2002,
    author = {Rosenberg, J\"org and Necker, Reinhold},
    doi = {10.1002/cne.10240},
    issn = {1096-9861},
    journal = {Journal of Comparative Neurology},
    number = {3},
    pages = {274-285},
    title = {Ultrastructural characterization of the accessory lobes of Lachi in the lumbosacral spinal cord of the pigeon with special reference to intrinsic mechanoreceptors},
    volume = {447},
    year = {2002}
}

@article{ruppert_learning_2021,
    author = {Ruppert, Felix and Badri-Spr{\"o}witz, Alexander},
    doi = {10.1038/s42256-022-00505-4},
    journal = {Nature Machine Intelligence},
    month = {November},
    number = {7},
    pages = {652--660},
    title = {Learning {Neuroplastic} {Matching} of {Robot} {Dynamics} in {Closed}-loop {CPGs}},
    url = {https://www.nature.com/articles/s42256-022-00505-4},
    volume = {4},
    year = {2022}
}

@article{schroeder_marginal_1986,
    author = {Schroeder, Dolores M},
    journal = {The Ohio Journal of Science},
    number = {3},
    pages = {69--72},
    title = {The marginal nuclei in the spinal cord of reptiles: intraspinal mechanoreceptors},
    volume = {86},
    year = {1986}
}

@article{schroeder_specializations_1987,
    author = {Schroeder, D. M. and Murray, R. G.},
    doi = {10.1002/jmor.1051940104},
    issn = {1097-4687},
    journal = {Journal of Morphology},
    number = {1},
    pages = {41--53},
    title = {Specializations within the lumbosacral spinal cord of the pigeon},
    url = {http://onlinelibrary.wiley.com/doi/10.1002/jmor.1051940104/abstract},
    volume = {194},
    year = {1987}
}

@article{Smith_2010,
    author = {Smith, Nicola C. and Jespers, Karin J. and Wilson, Alan M.},
    doi = {10.1242/jeb.020271},
    issn = {0022-0949},
    journal = {Journal of Experimental Biology},
    number = {8},
    pages = {1347-1355},
    title = {Ontogenetic scaling of locomotor kinetics and kinematics of the ostrich (Struthio camelus)},
    volume = {213},
    year = {2010}
}

@article{stanchak_balance_2020,
    author = {Stanchak, K E and French, C and Perkel, D J and Brunton, B W},
    doi = {10.1093/iob/obaa024},
    issn = {2517-4843},
    journal = {Integrative Organismal Biology},
    number = {obaa024},
    title = {The {Balance} {Hypothesis} for the {Avian} {Lumbosacral} {Organ} and an {Exploration} of {Its} {Morphological} {Variation}},
    url = {https://doi.org/10.1093/iob/obaa024},
    volume = {2},
    year = {2020}
}

@article{stanchak_molecular_2022,
    author = {Stanchak, Kathryn E. and Miller, Kimberly E. and Lumsden, Eric W. and Shikiar, Devany and Davis, Calvin and Brunton, Bingni W. and Perkel, David J.},
    doi = {10.1523/ENEURO.0100-22.2022},
    issn = {2373-2822},
    journal = {eNeuro},
    month = {August},
    pmid = {36008136},
    title = {Molecular markers of mechanosensation in glycinergic neurons in the avian lumbosacral spinal cord},
    url = {https://www.eneuro.org/content/early/2022/08/22/ENEURO.0100-22.2022},
    year = {2022}
}

@article{streeter1904structure,
    author = {Streeter, George L},
    doi = {10.1002/aja.1000030102},
    journal = {American Journal of Anatomy},
    number = {1},
    pages = {1--27},
    publisher = {Wiley Online Library},
    title = {The structure of the spinal cord of the ostrich},
    volume = {3},
    year = {1904}
}

@article{taga_emergence_1994,
    author = {Taga, Gentaro},
    doi = {10.1016/0167-2789(94)90283-6},
    issn = {0167-2789},
    journal = {Physica D: Nonlinear Phenomena},
    month = {August},
    number = {1-3},
    pages = {190--208},
    title = {Emergence of bipedal locomotion through entrainment among the neuro-musculo-skeletal system and the environment},
    url = {http://www.sciencedirect.com/science/article/pii/0167278994902836},
    volume = {75},
    year = {1994}
}

@article{tamura_variation_2007,
    author = {Tamura, Atsutaka and Nagayama, Kazuaki and Matsumoto, Takeo and Hayashi, Sadayuki},
    doi = {10.4271/2007-22-0006},
    journal = {Stapp car crash journal},
    month = {November},
    pages = {139--54},
    title = {Variation in nerve fiber strain in brain tissue subjected to uniaxial stretch},
    volume = {51},
    year = {2007}
}

@incollection{telano_physiology_2022,
    address = {Treasure Island (FL)},
    author = {Telano, Lauren N. and Baker, Stephen},
    pmid = {30085549},
    publisher = {StatPearls Publishing},
    title = {Physiology, {Cerebral} {Spinal} {Fluid}},
    url = {http://www.ncbi.nlm.nih.gov/books/NBK519007/},
    year = {2022}
}

@book{terni1924ricerche,
    author = {Terni, Tullio},
    publisher = {L. Niccolai},
    title = {Ricerche sulla cosidetta sostanza gelatinosa (corpo glicogenico) del midollo lombo-sacrale degli uccelli...},
    year = {1924}
}

@article{Thorson1973a,
    author = {Bied
erman-Thorson, Marguerite and Thorson, John},
    doi = {10.1007/BF00696890},
    issn = {1432-1351},
    journal = {Journal of comparative physiology},
    number = {2},
    pages = {103-122},
    title = {Rotation-compensating reflexes independent of the labyrinth and the eye},
    type = {Journal Article},
    url = {https://doi.org/10.1007/BF00696890
https://link.springer.com/content/pdf/10.1007/BF00696890.pdf},
    volume = {83},
    year = {1973}
}

@article{urbina2018physical,
    author = {Urbina-Mel{\'e}ndez, Dar{\'i}o and Jalaleddini, Kian and Daley, Monica A and Valero-Cuevas, Francisco J},
    doi = {10.3389/frobt.2018.00038},
    journal = {Frontiers in Robotics and AI},
    pages = {38},
    publisher = {Frontiers},
    title = {A physical model suggests that hip-localized balance sense in birds improves state estimation in perching: implications for bipedal robots},
    volume = {5},
    year = {2018}
}

@article{viana_di_prisco_synaptic_1990,
    author = {Viana Di Prisco, G. and Walle´n, P. and Grillner, S.},
    doi = {10.1016/0006-8993(90)90675-2},
    issn = {0006-8993},
    journal = {Brain Research},
    month = {October},
    number = {1},
    pages = {161--166},
    title = {Synaptic effects of intraspinal stretch receptor neurons mediating movement-related feedback during locomotion},
    url = {http://www.sciencedirect.com/science/article/pii/0006899390906752},
    volume = {530},
    year = {1990}
}

@article{watterson_development_1949,
    author = {Watterson, Ray L. and Spiroff, Boris E. N.},
    doi = {10.2307/30152058},
    issn = {0031-935X},
    journal = {Physiological Zoology},
    month = {October},
    number = {4},
    pages = {318--337},
    title = {Development of the {Glycogen} {Body} of the {Chick} {Spinal} {Cord}. {II}. {Effects} of {Unilateral} and {Bilateral} {Leg}-{Bud} {Extirpation}},
    url = {http://www.jstor.org/stable/30152058},
    volume = {22},
    year = {1949}
}

@article{wedel2021expanded,
    author = {Wedel, M and Atterholt, J and Dooley, A and Farooq, S and Macalino, J and Nalley, T and Wisser, G and Yasmer, J},
    doi = {10.20935/AL911},
    journal = {Acad. Lett},
    title = {Expanded neural canals in the caudal vertebrae of a specimen of Haplocanthosaurus},
    volume = {911},
    year = {2021}
}

@article{yamanaka2008chick,
    author = {Yamanaka, Yuko and Kitamura, Naoki and Shibuya, Izumi},
    doi = {10.2220/biomedres.29.205},
    journal = {Biomedical Research},
    number = {4},
    pages = {205--211},
    publisher = {Biomedical Research Press},
    title = {Chick spinal accessory lobes contain functional neurons expressing voltagegated sodium channels generating action potentials},
    volume = {29},
    year = {2008}
}

@article{yamanaka_analysis_2012,
    author = {Yamanaka, Yuko and Kitamura, Naoki and Shinohara, Hikaru and Takahashi, Keita and Shibuya, Izumi},
    doi = {10.1007/s00359-011-0703-0},
    journal = {Journal of Comparative Physiology A},
    number = {3},
    pages = {229--237},
    title = {Analysis of {GABA}-induced inhibition of spontaneous firing in chick accessory lobe neurons},
    volume = {198},
    year = {2012}
}

@article{yamanaka_glutamate_2013,
    author = {Yamanaka, Yuko and Kitamura, Naoki and Shinohara, Hikaru and Takahashi, Keita and Shibuya, Izumi},
    doi = {10.1007/s00359-012-0766-6},
    issn = {0340-7594, 1432-1351},
    journal = {Journal of Comparative Physiology A},
    month = {January},
    number = {1},
    pages = {35--43},
    title = {Glutamate evokes firing through activation of kainate receptors in chick accessory lobe neurons},
    url = {http://link.springer.com/article/10.1007/s00359-012-0766-6},
    volume = {199},
    year = {2013}
}

\end{document}